\documentclass[letter]{IEEEtran}

\usepackage{amsmath,amssymb,amsthm}
\usepackage{epsfig,cite,footnote,authblk,mathtools,enumerate}
\usepackage{color}
\usepackage[table]{xcolor}

\definecolor{bluecolor}{rgb}{0,0.,1.}

\definecolor{redcolor}{rgb}{.7,0.,0.}

\newcommand{\es}[1]{\begin{equation}\begin{split}#1\end{split}\end{equation}}

\newcommand{\V}{\mathcal{V}}

\newcommand{\BW}{\mathrm{BW}}

\newcommand{\dd}{\textrm{d}}

\newcommand{\SF}{\text{SF}}

\newcommand{\squeezeup}{\vspace{-1.3mm}}

\IEEEoverridecommandlockouts

\begin{document}

\title{\huge{Optimal Non-Uniform Deployments of LoRa Networks}}

\author{Orestis Georgiou \textit{Senior Member, IEEE}, Constantinos Psomas, \textit{Senior Member, IEEE}, Christodoulos Skouroumounis, \textit{Member, IEEE},  and Ioannis Krikidis \textit{Fellow, IEEE}

\thanks{All authors are with the IRIDA Research Centre for Communication Technologies, Department of Electrical and Computer Engineering, University of Cyprus, Nicosia, Cyprus (e-mail: \{georgiou.orestis\}@ucy.ac.cy).}

}

\maketitle

\bstctlcite{IEEEexample:BSTcontrol}

\vspace*{-1.1cm}
\begin{abstract}
LoRa wireless technology is an increasingly prominent solution for massive connectivity and the Internet of Things.  Stochastic geometry and numerical analysis of LoRa networks usually consider  uniform end-device deployments. Real deployments however will often be non-uniform, for example due to mobility. This letter mathematically investigates how non-uniform deployments affect network coverage and suggest optimal deployment strategies and uplink random access transmission schemes. We find that concave deployments of LoRa end-devices with a sub-linear spread of random access inter-transmission times provide optimal network coverage performance.
\end{abstract}



\squeezeup
\section{Introduction \label{sec:intro}}

Fuelled by the smart-city vision and Internet of Things massive connectivity applications, low power wide area network (LPWAN) technologies have recently seen a dramatic increase in academic research interest and industrial deployments \cite{raza2017low}.
Specifically, LoRa (Long Range) technology has emerged as an interesting solution for both urban and rural sensing and control applications (e.g., smart metering, agriculture, supply chain \& logistics) due to its attractive long range, low power, and low cost features as well as its ease of deployment and use of unlicensed radio spectrum \cite{sornin2015lorawan}.
It is thus presented as a good solution for moderately dense networks of low traffic devices, which do not impose strict latency or reliability requirements. 

To that end, many works have investigated the performance of LoRa networks, with particular interest on its scalability and interference management, with the aim to suggest simple engineering solutions that can improve reliability, scalability, and information delay \cite{beltramelli2018interference,lim2018spreading,ochoa2018large}. 
In this letter, we build on previous results \cite{georgiou2017low} and leverage tools from stochastic geometry \cite{haenggi2012stochastic}
to
 \textit{i)} present a novel mathematical framework for uplink performance analysis that models non-uniform deployments, 
\textit{ii)} derive expressions for the packet collision probability and coverage probability using meta distribution statistics of the inherent interference and demonstrate how the two are related, and finally \textit{iii)} define optimization problems used to obtain optimal and fair deployment strategies.

\squeezeup
\section{System Model \label{sec:system}}

\begin{figure}[t]
\centering
\includegraphics[width=\columnwidth]{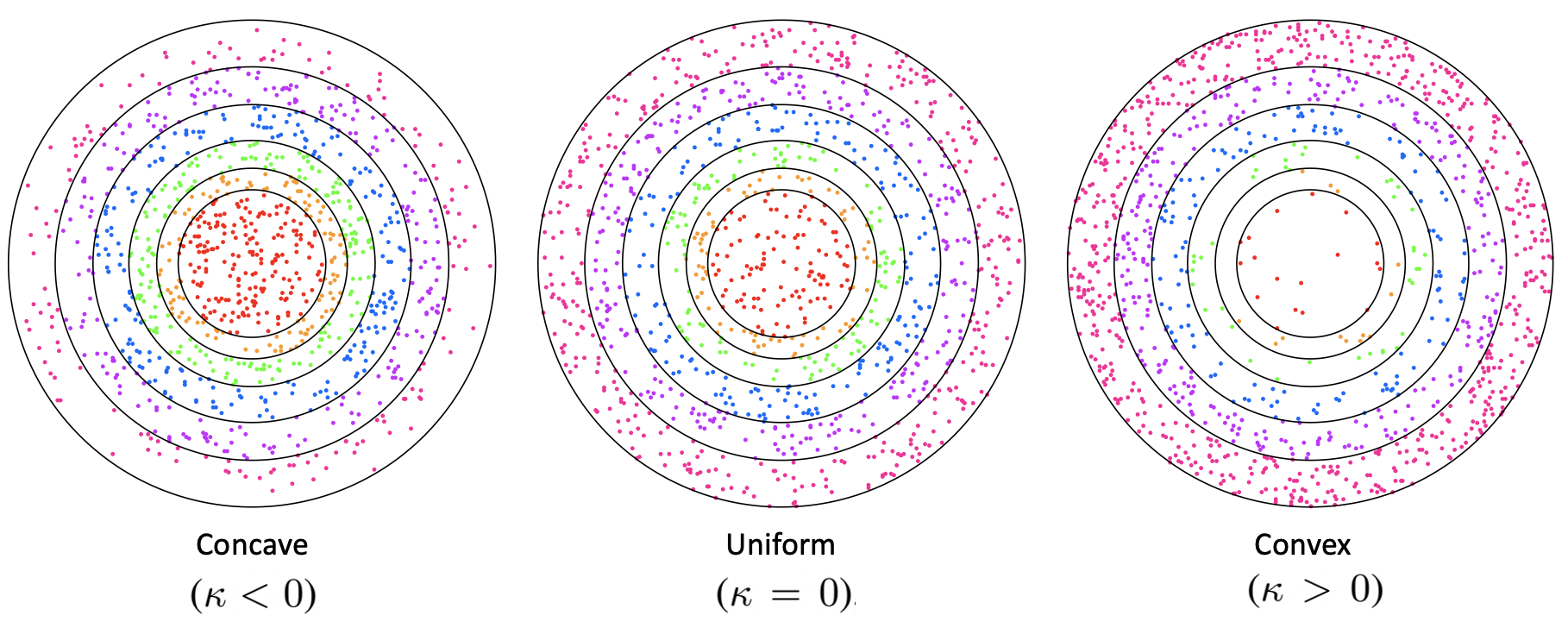}
\caption{ 
Non-uniform distributions of LoRa EDs using $\lambda_0=3$ EDs per km${}^2$. The SF rings are shown as concentric circles of radius $l_n$ km for $n\in[1,6]$.
}
\label{fig:nonuniform}
\end{figure}

We consider a LoRa network comprised of a central gateway (GW) and many end-devices (EDs) that are located within a circular deployment region $\V$ of radius $R$ km according to a inhomogeneous Poisson point process (PPP) $\Phi$ with intensity  $ \lambda(d_i)= \lambda_0 (1+ \kappa (d_i^2 - R^2/2)) $ where $d_i \in(0,R]$ km is the radial distance of ED $i$ from the GW, $\lambda_0 > 0$ is the average intensity of the PPP, and $\kappa\in[-\frac{2}{R^2}, \frac{2}{R^2}]$ is a curvature parameter such that there are, on average, $N=\lambda_0 |\mathcal{V}| = 2\pi \int_0^R \lambda(x)x  \dd x$ EDs in $\mathcal{V}$. 
Importantly, the curvature parameter $\kappa$ controls how the EDs are deployed within $\mathcal{V}$ and allows one to interpolate between uniform deployments ($\kappa=0$), concave deployments ($\kappa<0$) where nodes are deployed predominantly near the central GW, and convex deployments ($\kappa>0$) where nodes are deployed predominantly near the network edge as seen in Fig. \ref{fig:nonuniform}, with each point corresponding to an ED.
Importantly, note that the concave case of $\kappa=-\frac{2}{R^2}$ is akin to the  stationary distribution of the Random Waypoint Mobility Model (RWPM)  \cite{bettstetter2003node}. 

To ensure mathematical tractability we will assume a path loss attenuation function defined through 
$
g(d_i)= (\psi/4\pi d_i)^\eta
$, 
where $\psi=34.5$ cm is the carrier wavelength ($f= 915$ MHz), $d_i$ is the Euclidean distance in km between ED $i$ and the central GW, and $\eta\geq 2$ is the path loss exponent.
Also, all wireless links are assumed to be subjected to both small-scale block fading and large-scale path-loss effects. 
We consider Rayleigh fading, with no shadowing effects, such that the channel gain $|h_i|^2$ due to an uplink transmission by ED $i$ is modelled as an exponential random variable of unit mean.

LoRa networks employ a proprietary chirp spread spectrum (CSS) modulation scheme that supports adaptive data rates (ADRs), thus enabling the system to trade-off throughput for coverage range, or robustness, or energy consumption, while keeping a constant bandwidth $\text{BW}$. 
This process is  managed by the network server (NS) that regulates the transmission spreading factor (SF), which in turn determines the length of the chirp symbol $T_s = 2^\SF / \BW$, and the bit rate $R_n=\frac{4\SF}{(4+\text{CR})T_S} $, where the coding rate $\text{CR}\in[1,4]$. 
The SFs range from 7 to 12 and are allocated to each ED by the NS according to a signal-to-noise-ratio (SNR) link margin feedback received in response to short test frames sent out by each ED after joining a network \cite{sornin2015lorawan}.
Using the SNR thresholds provided by Semtech \cite{sornin2015lorawan}, we can thus model the SF assignment to EDs as a kind of tiered ring structure around each GW shown in Fig. \ref{fig:nonuniform}, in different colors. 
Thus, EDs within each ring are assigned a unique SF and SNR threshold $q_n$ based on their distance to the origin and the interval $(l_{n-1},l_n)$ this falls into, where $n\in[1,6]$.
We can set the outer radius of the $n_\text{th}$ ring $l_n$ to satisfy $\mathbb{E}[\text{SNR}]=10^{\frac{\mathcal{P-\mathcal{N}}}{10}}g(l_n) \mathbb{E}[|h_i|^2]\geq q_n$, where $\mathcal{P}$ is the transmit power in dBm, and $q_{n}= -6,-9,-12,-15,-17.5,-19$ dBm is the SNR threshold of the corresponding SF, $ \mathcal{N} = -174+ \text{NF} + 10\log\text{BW}$ dBm is the noise power, $\text{NF}=6$ dBm is the typical receiver noise figure, and we explicitly define $l_0=0$ and set $R=l_6$ km.
Re-arranging the inequality we have that 
$
l_n = \frac{\psi}{4 \pi} 10^{\frac{ \mathcal{P}-q_{n+6 - \mathcal{N}}}{10\eta}}
$
and using that the maximum LoRa uplink transmission power is limited to 25 mW ($\mathcal{P}\approx 14$ dBm) and a path loss exponent of $\eta=2.7$ to represent sub-urban environments we get
$l_n\approx 3.3, 4.2, 5.5, 7.0 , 8.7, 10.8$ km.

Unlike other communication systems, e.g., cellular, where interference is treated as shot-noise and link performance is measured by the signal-to-interference-plus-noise-ratio (SINR) condition \cite{haenggi2012stochastic}, LoRa uplink transmissions have a dual requirement that separates the SNR and signal-to-interference-ratio (SIR) \cite{georgiou2017low,sornin2015lorawan}.
This is due to the CSS modulation scheme which allows GWs to decode multiple received transmissions, as long as their relative signal strengths are sufficiently distinct \cite{beltramelli2018interference}.
Specifically, we say that an uplink transmission by ED $i$ is successful if the following condition is met
\es{
    \big(  \text{SNR}_{i} \geq q_n \big)   \bigcap \big( \text{SIR}_{i} \geq w \big)  ,
\label{SINR}
}
where we define 
$\text{SNR}_{i}= \frac{\mathcal{P}  |h_{i} |^2 g(d_i)}{\mathcal{N}}$ 
and 
$\text{SIR}_{i}= \frac{  \mathcal{P}  |h_{i} |^2 g(d_i)}{ \mathcal{I}_i  }
$, respectively,  $w= 1.259 = 1$ dB is the co-SF threshold, and  \cite{beltramelli2018interference}
$
\mathcal{I}_i=\sum_{k\not=i} \chi_{ik}  \mathcal{P}  |h_{k}|^2 g(d_{k})
$ 
is the total co-SF interference at the GW, with  $\chi_{ik}=1$ if ED $k$ is transmitting with the same SF as ED $i$, and zero otherwise, the average of which is the packet collision probability  $\mathbb{E}[\chi_{ik}] = p(d_i)$. 
Inter-SF interference is not considered in this paper.

\squeezeup
\section{Packet Collision Probability \label{sec:packet}}

\begin{figure}[t]
\centering
\includegraphics[scale=0.25]{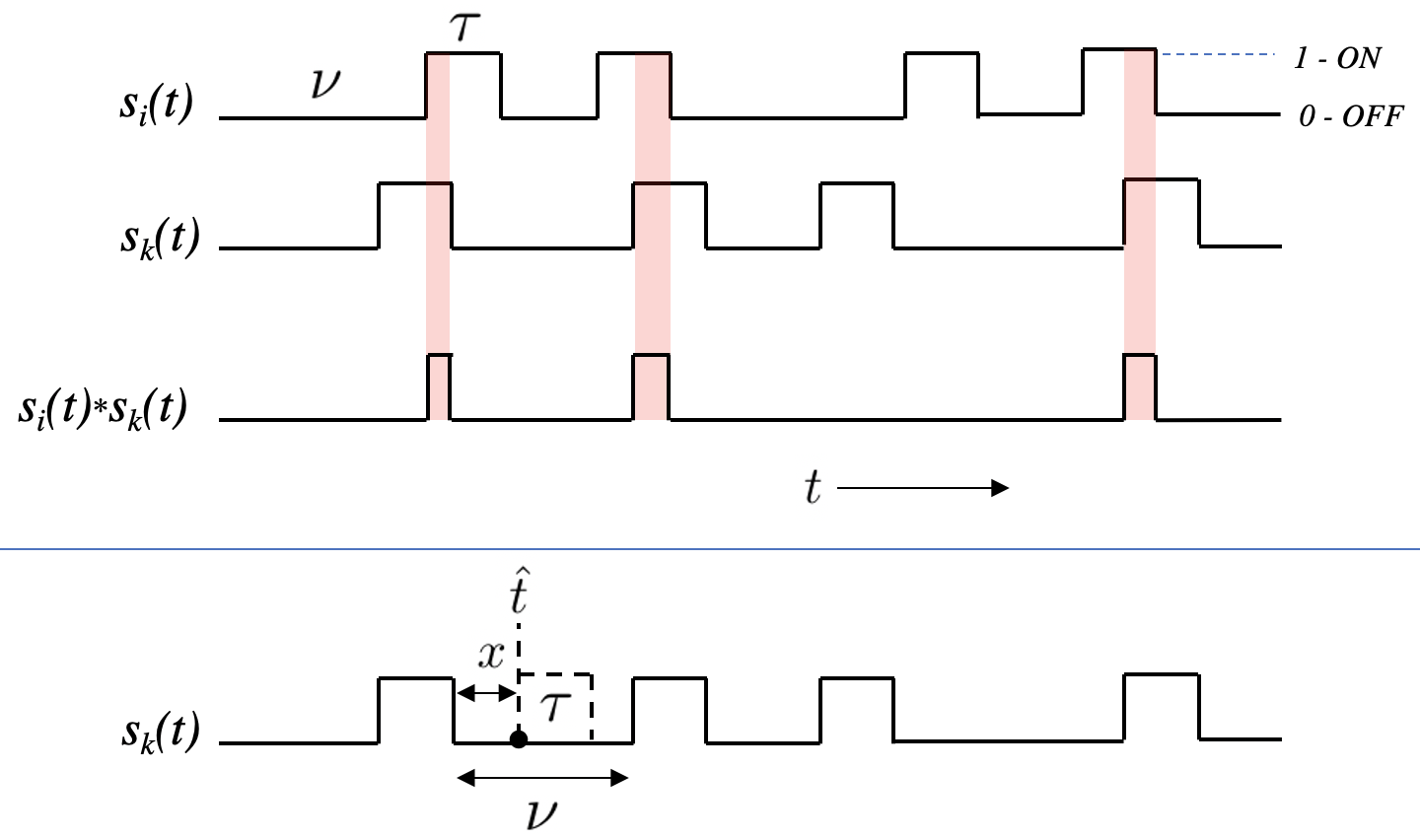}
\caption{ 
\textit{Top:} Two asynchronous transmission streams $s_i(t)$ and $s_k(t)$ and their overlap $s_i(t)s_k(t)$ showing the collision intervals. The two streams are basically similar to unslotted ALOHA transmissions, but with independent and identically distributed packet inter-arrival times given by the random variable $\nu$, and a constant packet air-time $\tau$ that depends on the SF used.
\textit{Bottom:} Equivalent representation of a packet collision. A point $\hat{t}$ is chosen at random during one of the non-transmitting sub-intervals of stream $s_k(t)$, with $x\in(0,\nu)$ being the time since the end of the previous transmission. A collision will not occur if $x+\tau <\nu$.
}
\label{fig:collisions}
\end{figure}

\begin{figure}[t]
\centering
\includegraphics[width=0.85\columnwidth]{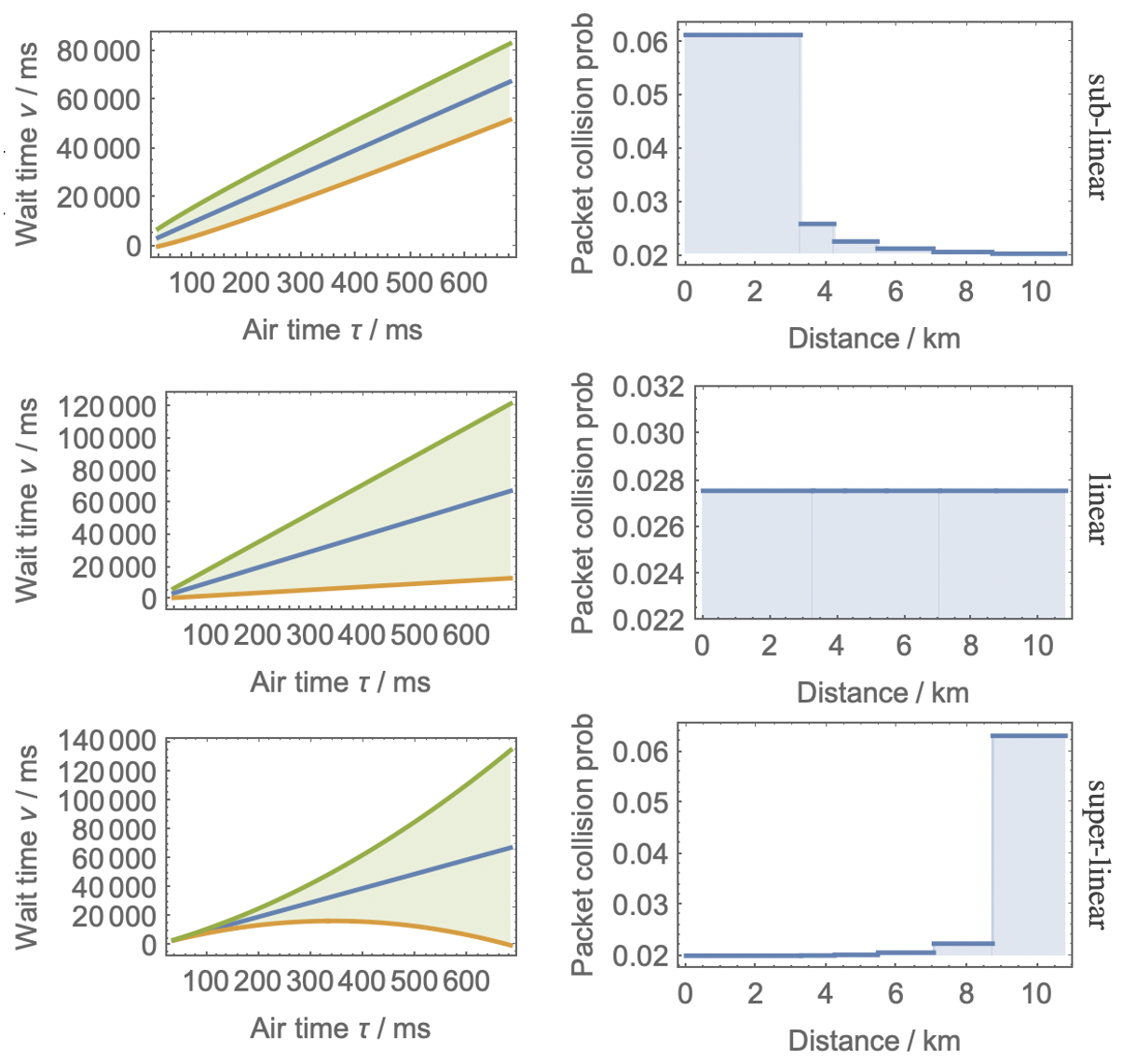}
\caption{ 
Air times vs. inter-transmission times and how they affect the packet collision probability $p(d_i)$ using $u=99$, $v(\tau_i)=598 \sqrt{\tau_i}$ (sub-linear), $v(\tau_i)=80 \tau_i$ (linear) and $v(\tau_i)=0.145 \tau_i^2$ (super-linear).
In all cases, the mean inter-transmission time is proportional to the transmission time $u\tau_i$ (blue straight line) taking values between $u\tau_i\pm v(\tau_i)$ (green and orange curves), while the corresponding packet collision probability goes from 2\% to 6\% depending on the SF used by the ED at $d_i$ km. 
From \cite[Tab. I]{georgiou2017low}, we have that the air times for each SF are $\tau_i=36.6, 64, 113, 204, 372, 682$ ms. 
These are used to calculate $\nu_1$ and $\nu_2$, and thus $p(d_i)$ is plotted on the right for the chosen $u$ and $v(\tau_i)$.
}
\label{fig:packet}
\end{figure}

LoRa EDs  employ the LoRaWAN MAC protocol wherein uplink transmissions are initiated by EDs with a small variation based on a random time basis; an ALOHA-like protocol \cite{sornin2015lorawan}.
We can therefore model the uplink transmission stream as a sequence of inter-transmission times $\nu_i$ ms and air-times $\tau_i$ ms that depend on the SF used by ED $i$ and are interrelated
by the European Telecommunications Standards Institute (ETSI) imposed duty cycle condition requiring that $\mathbb{E}[\frac{\tau_i}{\nu_i+\tau_i}]\leq  1\%$.
For simplicity, we take $\tau_i$ to be equal to the packet payload size (in bits) divided by the corresponding bit-rate $R_n$ (i.e., we ignore the time needed to transmit the preamble and any variable coding rate effects), and $\nu_i$ is a uniformly distributed random variable with probability density function (pdf) given by $
f_{\nu_i}(x) = \frac{1}{\nu_2-\nu_1}$ with $ x\in[\nu_1, \nu_2]$.
Bit-rates and air times $\tau_i$ for a 25 byte message are provided  in \cite[Tab. I]{georgiou2017low} and are used in Figs. \ref{fig:packet}-\ref{fig:optimal}. 
The 1\% duty cycle condition implies that higher SFs that have longer transmission times will necessarily have longer inter-transmissions times, on average.

To calculate the packet collision probability $p(d_i)$ of two asynchronous transmission streams $s_i(t)$ and $s_k(t)$ of the same SF (see Fig. \ref{fig:collisions}) one could numerically simulate
$
p(d_i) = \lim_{T\to\infty}\frac{1}{T}\int_0^T s_i(t)s_k(t)\dd t
$
which gives the time averaged portion of time during which the two co-SF streams overlap.  
An equivalent way of looking at this picture is shown in the bottom sub-figure of Fig. \ref{fig:collisions}, where we choose a  random start point of the transmission by ED $i$ denoted by $\hat{t}$ relative to the transmission stream $s_k(t)$ of the interfering ED $k$, and write  
\es{
&p(d_i)= 1- \mathbb{P}[\text{ED $k$ is off at $t\!=\!\hat{t}$ \& stays off until $t\!=\!\hat{t}\!+\!\tau_i$}]\\ 
&=1-\mathbb{E}_{\nu}\Big[\frac{\nu_i}{\tau_i +\nu_i}\Big] \times \mathbb{E}_{\nu_i}\big[\mathbb{P}[X+\tau_i<\nu_i \big| \,\nu_i]\big]\\ 
&=1-\Big(\!\int_{\nu_1}^{\nu_2}\! \frac{x}{\tau_i +x} f_{\nu_i}(x)\dd x \Big) \Big(\!\int_{\hat{\nu}_1}^{\nu_2}\!\! \int_0^{\nu_i-\tau_i}\!\!\! f_X(x) \dd x \,\dd \nu_i \Big) ,
\label{pdi}}
where in the second line we identify the first term in the product as the complement of duty cycle condition which should be greater than 99\%.
The second term in the product is conditional on the inter-transmission time $\nu_i$ so as to first calculate the probability that the time $X$ since the last uplink transmission of ED $k$ ended, denoted by $X=x$ in Fig. \ref{fig:collisions}, plus the air-time $\tau_i$ of node $i$ is less than $\nu_i$.
Note that the lower integration limit for $\nu$ is  $\hat{\nu}_1=\text{max}(\nu_1,\tau_i)$, since we require that $0<x<\nu_i-\tau_i$.
Since $\hat{t}$ was chosen at random within an inter-transmission time interval of ED $k$, it follows that $X$ is uniform in $(0,\nu)$, i.e., $f_X(x)=\frac{1}{\nu}$, giving
\es{
p(d_i)&=1 -  \Big(1 -\frac{\tau_i}{\nu_2 - \nu_1} \ln \frac{\nu_2+\tau_i}{\nu_1+\tau_i}\Big)\Big( \frac{\nu_2-\hat{\nu}_1- \tau_i \ln \frac{\nu_2}{\hat{\nu}_1}}{\nu_2 - \nu_1} \Big)  .
\label{colU}
}
Note that the inter-transmission time $\nu_i$ and transmission time $\tau_i$ are interrelated via the values of $\nu_1$ and $\nu_2$ for each SF that are chosen such that the mean inter-transmission time satisfies $\mathbb{E}[\nu_i]=(\nu_2+\nu_1)/2 \geq  99\tau_i$ and is thus aligned with the ETSI duty cycle condition of 1\%. 
We therefore  define $\nu_2= u\tau_i + v(\tau_i)$ and $\nu_1= u\tau_i- v(\tau_i)$ with $u\geq 99$ such that the mean inter-transmission time is always $\mathbb{E}[\nu_i]= u\tau_i$,
and the function $v(\tau_i)$ controls its variance $\text{var}(\nu_i)=v(\tau_i)^2/3$ and higher moments.

We will now show that $v(\tau_i)$ can be chosen to influence the packet collision probability $p(d_i)$ through \eqref{colU}.
For example, a linear $v(\tau_i) = c \tau_i$  results in a constant $p(d_i)= \frac{1}{c}\ln\frac{u+c}{u-c}$ that does not depend on $\tau_i$ and therefore does not depend on the ED $i$ location $d_i$.
Instead, any super(sub)-linear function $v(\tau_i)$ will result in an increasing(decreasing) $p(d_i)$.
For example, setting $v(\tau_i)= c\tau_i \ln \tau_i $ gives $p(d_i)\approx \frac{1}{c \ln \tau}\ln\frac{u+c\ln \tau_i}{u-c\ln\tau_i}$, while setting $v(\tau_i)= c\tau_i / \ln \tau_i $ gives $p(d_i)\approx \frac{\ln\tau_i}{c }\ln\frac{u\ln \tau_i+c}{u\ln\tau_i-c}$.
Similarly, using simpler functions such as a quadratic $v(\tau_i)= c\tau_i^2 $ or a square root $v(\tau_i)= c\sqrt{\tau_i} $ function gives $p(d_i)\approx \frac{2}{u} + \frac{2c^2 \tau_i^2}{3u^3}$ and $p(d_i)\approx \frac{2}{u} + \frac{2c^2}{3\tau_i u^3}$, respectively. 
These approximations are obtained by Taylor expanding \eqref{colU} with respect to $\tau_i$ and using the chosen $v(\tau_i)$ scalings of $\nu_1$ and $\nu_2$.
It follows, that one can influence network performance with respect to the SIR condition \eqref{laplace} via $p(d_i)$ by engineering $v(\tau_i)$ (see Fig. \ref{fig:packet}), a novel design function that has never been studied before.

\squeezeup
\section{Uplink Outage Probability \label{sec:outage}}

\begin{figure}[t]
\centering
\includegraphics[width=\columnwidth]{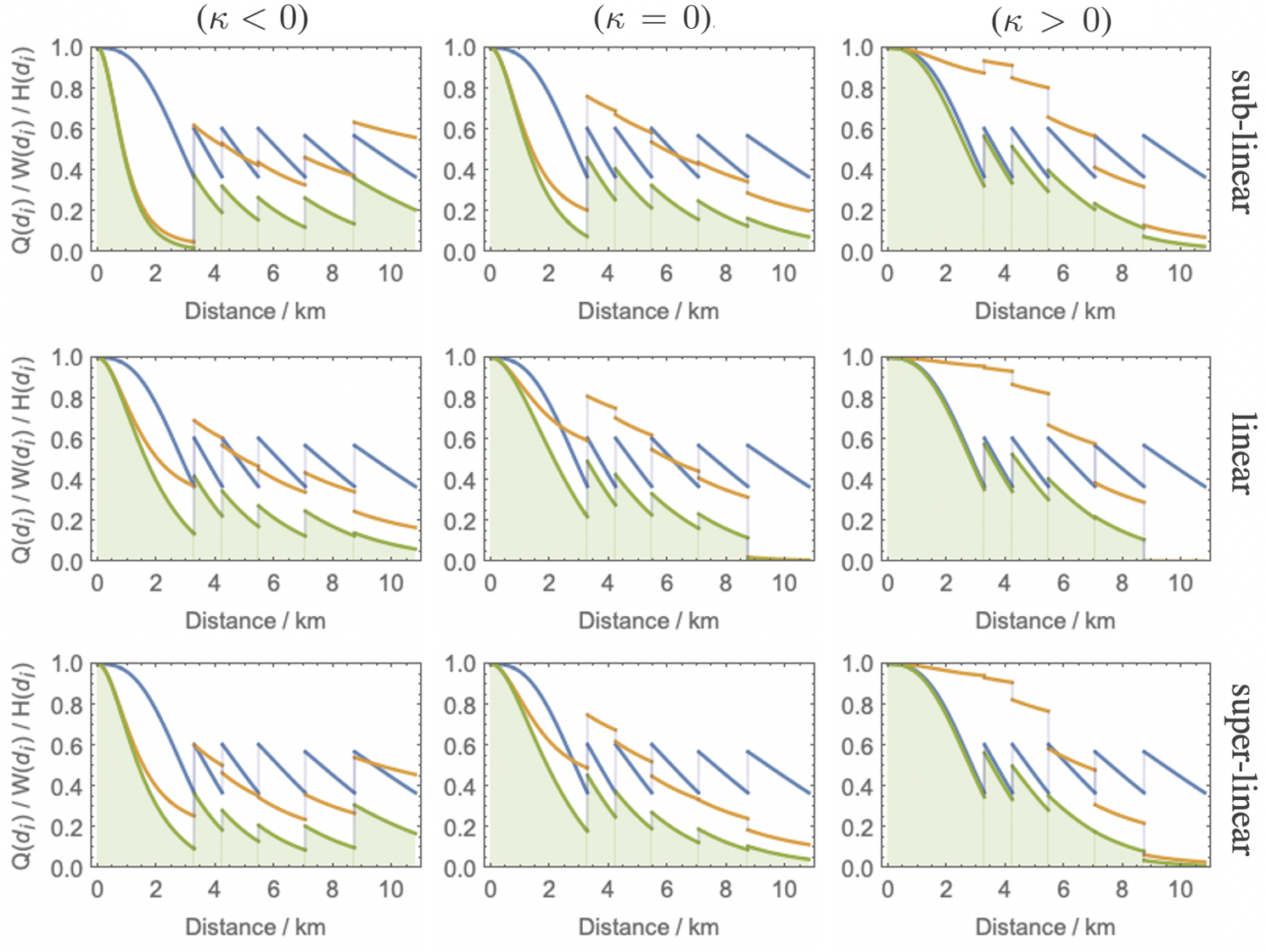}
\caption{ 
Plots of the SNR and SIR conditions $Q(d_i)$ $W(d_i)$ as well as the lower bound of $H(d_i)\geq Q(d_i)W(d_i)$ (blue, orange, and green for different values of the ED distribution curvature parameter $\kappa=-\frac{2}{R^2}, 0 ,\frac{2}{R^2}$, and for different inter-transmission time interval functions $v(\tau_i)$ (as those used in Fig. \ref{fig:packet}) and thus different packet collision probabilities $p(d_i)$. While the SNR condition $Q(d_i)$ is unaffected by $\kappa$ and $v(\tau_i)$, the SIR condition $W(d_i)$, and thus the uplink transmission success probability $H(d_i)$ are both strongly influenced by the network deployment and the uplink transmission scheme.
}
\label{fig:SINR}
\end{figure}

Equation \eqref{SINR}  captures the joint SNR and SIR decoding requirement at the GW, such that  the success probability of an uplink transmission by ED $i$ is 
$
H(d_i) \!=\! \mathbb{P} \big[   \text{SNR}_{i} \geq q_{n }  \bigcap \text{SIR}_{i} \geq w \big|\, d_i \big] \geq 
 Q(d_i) W(d_i)
$,
where we define $Q(d_i)=\mathbb{P} \big[    \text{SNR}_{i} \geq q_{n }  \,\big|\, d_i \big]$ and $W(d_i)=
\mathbb{P} \big[  \text{SIR}_{i} \geq w \,\big|\, d_i \big]$ and assume independence between these events, thus resulting in the lower bound probability inequality (\cite[eq.(5)]{lim2018spreading}). 
From the definition of SNR we  immediately have  
$
Q(d_i) =  \mathbb{P} \big[    |h_{i}|^2 \geq \frac{\mathcal{N} q_{n }}{\mathcal{P}g(d_{i})} \,\big|\, d_i \big] = e^{- \frac{\mathcal{N} q_{n}}{\mathcal{P}g(d_{i})}} $,
and note that $q_{n }$ is essentially a piecewise constant function of distance $d_{i}$, the effect of which is clearly seen in Fig. \ref{fig:SINR}.
Similarly for SIR 
$
W(d_i)= \mathbb{E}_{\mathcal{I}_i}\Big[\mathbb{P} \big[  |h_{i}|^2 \geq \frac{w \mathcal{I}_i}{\mathcal{P} g(d_{i})}  \,\Big|\, \mathcal{I}_i\, ,\, d_i \big]\Big] 
= \mathbb{E}_{\mathcal{I}_i} \big[ e^{-\frac{w \mathcal{I}_i}{\mathcal{P} g(d_{i})}  }\big] ,
$,
where the last term is the Laplace transform of the random variable $\mathcal{I}_i$ evaluated
at $ \frac{w }{\mathcal{P} g(d_{i})}$ conditioned on the location of  ED $i$ at $d_i$.
Note that an upper bound to $H(d_i)$ can be obtained by using $\max(x,y)=(x+y+|x-y|)/2$ and realizing that $H(d_i)=\mathbb{P}\big[|h_i|^2\geq  \max \big(\frac{\mathcal{N} q_{n }}{\mathcal{P}g(d_{i})}, \frac{w\mathcal{I}_i}{\mathcal{P}g(d_{i})} \big) \big| \, d_i \big]
\leq e^{- \frac{\mathcal{N} q_{n}}{2\mathcal{P}g(d_{i})}}  \mathbb{E}_{\mathcal{I}_i} \Big[ e^{-\frac{w \mathcal{I}_i}{2\mathcal{P} g(d_{i})}  }\Big] $ which can be obtained from the closed form expressions for $Q(d_i)$ and $W(d_i)$ given in \eqref{MB2}. 
Also note that $H(d_i)$ approaches its upper bound when $\mathcal{N}q_n \approx w \mathcal{I}_i$, and the lower bound otherwise, i.e., when the network is either interference or noise-limited.

Expanding the Laplace transform of the interference we get
\es{
 \mathbb{E}_{d_{k}} \!\Big[\! \prod_{k\not=i}\! \mathbb{E}_{|h_{k}|^2} \! \Big[ e^{-   w \chi_{ik} |h_{k}|^2 \frac{d_{i}^\eta}{d_{k}^\eta} } \Big] \! \Big] 
\! = \! \mathbb{E}_{d_{k}} \! \Big[ \! \prod_{k\not= i} \! \frac{1}{1\!+\! w \chi_{ik}  \frac{d_{i}^\eta}{d_{k}^\eta}} \!\Big] ,
\label{SIR2}
}
since the channel gains $|h_{k}|^2$ were assumed independent exponential random variables and we used the fact that $\frac{g(a)}{g(b)}=(b/a)^\eta$ and $\mathbb{E}_{|h|^2} [e^{-x |h|^2}] = \int_0^\infty e^{-z(x+1)} \dd z= \frac{1}{1+x} $. 

Next, we would like to capture the average packet collision rate via a thinning process. 
Assuming that the co-SF interfering signals arriving at the GW originate from EDs that are spatially distributed according to a thinned inhomogeneous PPP with intensity $\hat{\lambda}(d)= p(d) \lambda(d)$ enables us to absorb the $\chi_{ik}$ term into $p(d_i)$ and use the probability generating functional (PGF) \cite{haenggi2012stochastic} of an inhomogeneous PPP 
$
\mathbb{E}\Big[\prod_{x\in\Phi } f(x) \Big] = \exp\Big(\int_{\V} (f(x)-1)\hat{\lambda}(x) \dd x \Big)
$ 
to arrive at 
\es{
W(d_i)= \exp\Big(\!\! -2 \pi p(d_i) \!\!  \int_{l_{n-1}}^{l_n}\!\!  \frac{w  \frac{d_{i}^\eta}{d_{k}^\eta}}{1+w  \frac{d_{i}^\eta}{d_{k}^\eta}} \lambda(d_k) d_{k}  \dd d_{k} \Big) ,
\label{laplace}
}
where $p(d_i)\in[0,1]$ is the packet collision probability at the GW of an uplink transmission from ED $i$ located at $d_i$.
The integral in \eqref{laplace} will be given in closed form in \eqref{MB2} Sec. \ref{sec:meta} when we consider generalizations of $W(d_i)$.

Fig. \ref{fig:SINR} provides a visual comparison matrix of three performance metrics ($Q/W/H$) for three different deployment strategies (convex/uniform/concave) and three different inter-transmission time interval functions. 
A great variability is observed, particularly in SF7 and SF12.
Thus, our derived analytical expressions can provide engineering insights regarding network deployment and transmission schemes.

\squeezeup
\section{Coverage Probability \label{sec:coverage}}

\begin{figure}[t]
\centering
\includegraphics[width=\columnwidth]{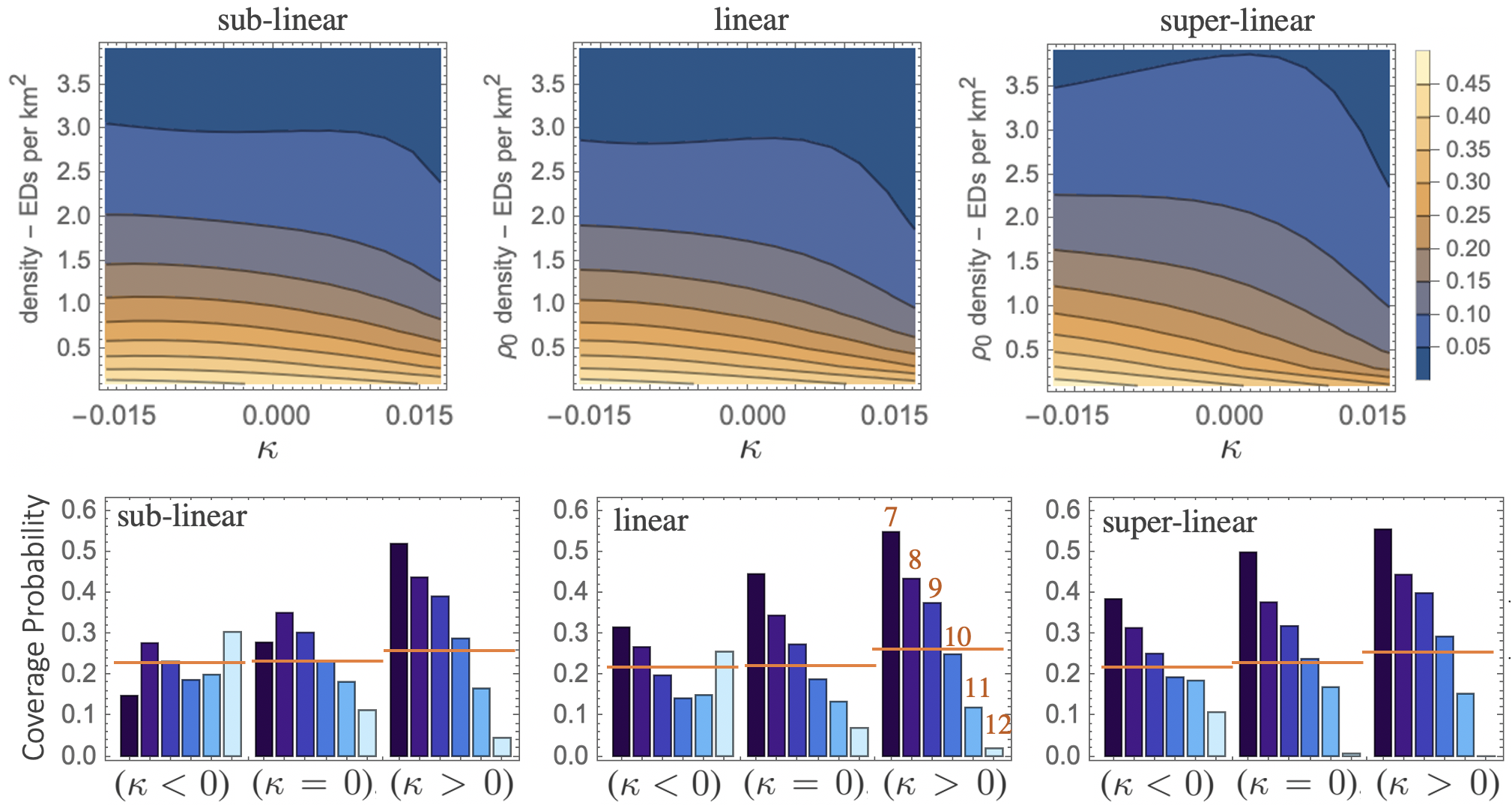}
\caption{ 
\textit{Top:} Contour plot of the coverage probability $C(\kappa,\lambda_0)$ for the three inter-transmission time functions $v(\tau)$ (as those used in Fig. \ref{fig:packet}).
\textit{Bottom:} Bar charts of the per-SF coverage probability $C_n(\kappa,\lambda_0)$  for different values of the ED distribution parameter $\kappa\!=\!-\frac{2}{R^2}, 0 ,\frac{2}{R^2}$, $\lambda_0\!=\!1$, and different $v(\tau)$ as above.
The horizontal  lines depict $\frac{1}{6}\sum_{n=1}^6\! C_n(\kappa,\lambda_0)$ for each condition. 
}
\label{fig:coverage}
\end{figure}

The coverage probability is the probability that a randomly selected ED is in coverage (i.e., not in outage) at any particular instance of time. 
One may obtain the system's coverage probability $C(\kappa,\lambda_0)$ by de-conditioning on the position of the specific ED $i$ achieved by averaging $H(d_i)$ over the deployment region $\V$ according to the PPP density. We will work with the lower bound of $H(d_i)\geq Q(d_i)W(d_i)$  since this is more relevant towards the performance analysis of LoRa
\es{
C(\kappa,\lambda_0) 
 =   \frac{2\pi}{N} \int_0^R Q(d_i)W(d_i) \lambda(d_i) d_i \dd d_i .
\label{C}}
However, despite having averaged over space $d_i$, the effective coverage probability is expected to be highly position dependent, especially in LoRa networks where different SNR thresholds $q_n$ apply in each SF ring.
We thus define a more granular per-SF coverage probability given by
\es{
C_n(\kappa,\lambda_0) &= \frac{2\pi }{N_n} \int_{l_{n-1}}^{l_n} Q(d_i)W(d_i) \lambda(d_i) d_i \dd d_i ,
\label{Cn}}
where we use $N_n= 2\pi \int_{l_{n-1}}^{l_n} \lambda(x) \dd x$ to indicate the average number of EDs in the $n_\text{th}$ SF ring. 
Note that $C(\kappa,\lambda_0)\not= \frac{1}{6}\sum_{n=1}^6 C_n(\kappa,\lambda_0)$.
Both coverage metrics \eqref{C} and \eqref{Cn} have been computed numerically and plotted in Fig. \ref{fig:coverage}.
First, observe that $C(\kappa,\lambda_0)$ shows a clear dependence on $\kappa$, $\lambda_0$, and the inter-transmission time statistics $v(\tau_i)$.
Second, observe that the distribution of per-SF coverage probabilities $C_n(\kappa,\lambda_0)$ varies significantly between SF rings, and is strongly dependent on $\kappa$ and $v(\tau)$; this is most noticeable in the case of $\SF =12$ where it can vary from $C_6(-\frac{2}{R^2},1)=0.3$ to $C_6(\frac{2}{R^2},1)=0$.
 

\squeezeup
\section{Coverage Meta Distribution \label{sec:meta}}

The coverage meta distribution (MD) has been
introduced as a performance metric that provides a more complete
spatial distribution rather than merely spatial averages as performed in the previous sections \cite{haenggi2015meta}. 
The MD relates to the conditional success probability of the SIR and is thus a  two-parameter distribution function defined by
$
\bar{F}(d_i,z)= \mathbb{P} \big[  W_\Phi(d_i) \geq z  \big]
$
where $z\in[0,1]$ is referred to as the reliability parameter and $W_\Phi(d_i)=\mathbb{P}[\text{SIR}_i \geq w \, \big|\,  \Phi ]$ is treated as a random variable conditioned on the spatial realization $\Phi$ of the PPP but averaged over the channel fading $|h_k|^2$.
The spatially averaged probability $W(d_i)$  can be retrieved from the MD directly from its definition
$
W(d_i)=\mathbb{E}_\Phi[W_\Phi(d_i)] = \int_0^1 \bar{F}(d_i,z) \dd z
$.
In practical terms, deconditioning the MD with respect to $d_i$ provides  the fraction of links whose SIR is greater than $w$ and their SNR is greater than the appropriate threshold $q_n$, with probability at least equal to $z$ in each PPP network realization:
\es{
\mathcal{C}(\kappa,\lambda_0,z)= \frac{2\pi}{N}\int_0^R Q(d_i)\bar{F}(d_i,z) \lambda(d_i)d_i\dd d_i ,
}
therefore yielding a more general statistical characterization of the performance of  LoRa networks.
Note that $C(\kappa,\lambda_0) = \int_0^1 \mathcal{C}(\kappa,\lambda_0,z) \dd z$, and unlike the standard coverage probability $C(\kappa,\lambda_0)$, the MD coverage probability $\mathcal{C}(\kappa,\lambda_0,z)$ answers some key questions network operators typically have, such as: ``What is the fraction of
EDs in a LoRa network that will achieve $z=90\%$ link reliability given
a deployment strategy of $\lambda(d)$?" or ``How will a different deployment strategy $\kappa$ affect the link reliability distribution among EDs?"

\begin{figure}[t]
\centering
\includegraphics[width=\columnwidth]{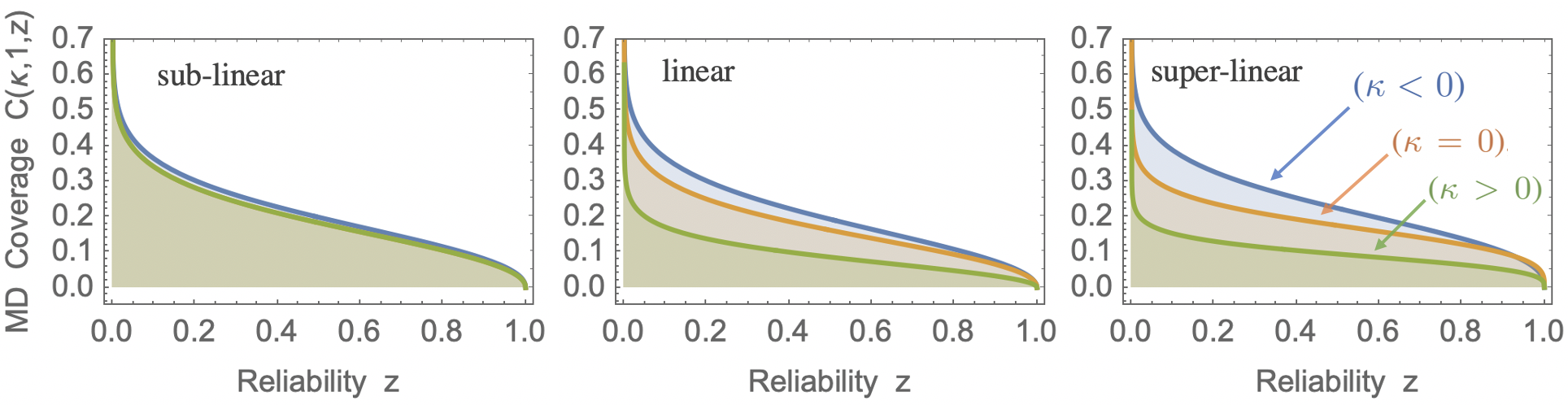}
\caption{ 
Plots of the MD coverage probability $\mathcal{C}(\kappa,\lambda_0=1,z)$ approximated by $1-\mathcal{G}(z)$ for different values of $\kappa$ and functions $v(\tau)$ as in previous figures.
}
\label{fig:MD}
\end{figure}

Calculating the MD $\bar{F}(d_i,z)$ is not straight forward and in most cases relies on obtaining the moments of $W_\Phi(d_i)$. 
One approach is to invoke the Gil-Pelaez inversion theorem, another is  to use the Fourier-Jacobi expansion, or  Mnatsakanov’s formula \cite{mnatsakanov2008hausdorff}.
Alternatively, one can approximate the MD by a Beta distribution with matched first and second moments since it has been shown to be an excellent approximation to $\bar{F}(d_i,z)$ \cite{haenggi2015meta}.
We thus define the $b_\text{th}$ moment of $W_\Phi(d_i)$ as
$M_b(d_i) = \mathbb{E}_\Phi[W_\Phi(d_i)^{b}] = \int_0^1 b z^{b-1} \bar{F}(d_i,z) \dd z$ 
and calculate
\es{
&M_b(d_i) = \mathbb{E}_\Phi[W_\Phi(d_i)^{b}] =  \mathbb{E}_{d_k} \Big[  \prod_{k\not= i}  \Big(1+ w \chi_{ik}  \frac{d_{i}^\eta}{d_{k}^\eta}\Big)^{-b} \Big] \\
&= \exp\!\Big( \! -2\pi p(d_i)   \int_{l_{n-1}}^{l_n} \!\! \Big(  1-  \Big(1+ w  \frac{d_{i}^\eta}{d_{k}^\eta}\Big)^{-b} \Big)\lambda(d_k) d_k \dd d_k ,
\label{MB}}
where we have followed a similar approach as in \eqref{SIR2} and used the PGF of a PPP.
Note that the first moment $M_1(d_i)= W(d_i)$  equals the success probability of the SIR condition \eqref{laplace}.
Performing the integral in \eqref{MB} we finally arrive at
\es{
M_b(d_i) &= \exp\bigg(\!-\frac{\pi p(d_i) \lambda_0}{2} \bigg[ 2 x^2 + \kappa (x-R)(x+R) \\
&+ x^2(\kappa R^2-2) {}_2F_1\Big(b,-\frac{2}{\eta},1-\frac{2}{\eta}, -w\frac{d_i^\eta}{x^\eta} \Big)\\
&- \kappa x^4  {}_2F_1\Big(b,-\frac{4}{\eta},1-\frac{4}{\eta}, -w\frac{d_i^\eta}{x^\eta} \Big) \bigg]_{x=l_{n-1}}^{x=l_{n}}  \bigg),
\label{MB2}}
where ${}_2F_1$ is the Gauss Hypergeometric function and can be calculated  to machine precision using standard software packages.
Averaging \eqref{MB2} over the PPP density multiplied by the SNR condition we obtain the coverage MD moments
$
\mathcal{M}_b = \frac{2\pi}{N}\int_0^R Q(d_i)M_b(d_i) \lambda(d_i)d_i\dd d_i $.
The Beta distribution has cumulative distribution function (cdf) $\mathcal{G}(x) = \frac{B(x;\alpha,\beta)}{B(\alpha,\beta)}
$,
where $B(x;\alpha,\beta)$ and $B(\alpha,\beta)$ are the incomplete and complete Beta functions, and the first two moments of which are
$
m_1=\frac{\alpha}{\alpha+\beta}$ and $ m_2=\frac{\alpha}{\alpha+\beta}\frac{\alpha+1}{\alpha+\beta+1} 
$, respectively, 
with $\alpha,\beta>0$.
By matching the first two moments of the Beta distribution with $\mathcal{M}_1$ and $\mathcal{M}_2$ and solving we get 
$
\alpha=\frac{\mathcal{M}_1(\mathcal{M}_1-\mathcal{M}_2)}{\mathcal{M}_2-\mathcal{M}_1^2}$, and $\beta
=\frac{(1-\mathcal{M}_1)(\mathcal{M}_1-\mathcal{M}_2)}{\mathcal{M}_2-\mathcal{M}_1^2}$,
thus allowing us to approximate $
\mathcal{C}(\kappa,\lambda_0,z)$ by $ 1-\mathcal{G}(z) $ with $\alpha$ and $\beta$ as above (see Fig. \ref{fig:MD}).
Note that $\mathcal{M}_2-\mathcal{M}_1^2$ corresponds to the variance of the meta coverage probability thus somewhat quantifying the differences (i.e., a measure of fairness) amongst the uplink per-SF coverage probability.
Observe in Fig. \ref{fig:MD} that $\mathcal{C}(\kappa,\lambda_0,z)$ is impervious to changes in ED deployment controlled through $\kappa$ for sub-linear inter-transmission times $v(\tau)$ but can vary greatly for linear and super-linear inter-transmission times demonstrating the fragility of EDs operating at higher SFs (i.e., long-distance uplink transmissions).
Moreover, note that the $\mathcal{C}(\kappa,\lambda_0,z)$ at $z=0.95$  represents the performance of the most susceptible EDs while  $z=0.05$  that of the best performing EDs. It does not however reveal the performance of EDs within each SF ring. 
To capture this variability, we  generalize the MD coverage moments through
\es{
\mathcal{M}_b^{(n)} = \frac{2\pi}{N_n} \int_{l_{n-1}}^{l_n}  Q(d_i)M_b(d_i) \lambda(d_i)d_i\dd d_i
}
and use these to calculate analogues of  $\alpha^{(n)}$, $\beta^{(n)}$ and thus obtain a per-SF MD of the coverage probability $\mathcal{C}_n(\kappa,\lambda_0,z)$ through a per-SF Beta  approximation $1-\mathcal{G}_n(z)$.
In this case, $ \mathcal{M}_{-1}^{(n)}$ can provide uplink delay insights such as the average number attempts before a successful transmission.

\squeezeup
\section{Coverage Optimization \label{sec:optimal}}

\begin{figure}[t]
\centering
\includegraphics[width=0.9\columnwidth]{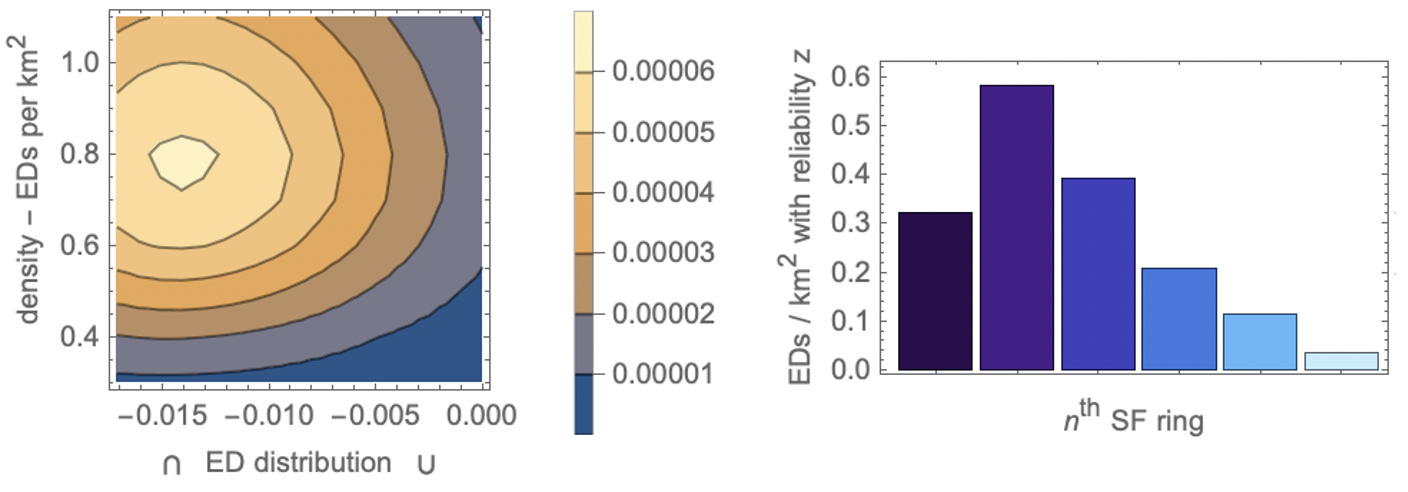}
\caption{\textit{Left:} Contour plot of the objective function $\mathcal{O}(\kappa,\lambda_0, 0.7)$.
\textit{Right:} Corresponding bar chart of the per-SF effective density $\mathcal{O}_n(-0.015,0.8, 0.7)$. 
}
\label{fig:optimal}
\end{figure}

Through our prior analysis, it transpires that there are many different performance metrics and many possible optimization avenues.
Namely, fusing the tools and formulation described herein, one may optimize with respect to the SF allocation rings defined through the radii $l_n\in(0,R)$ \cite{lim2018spreading}, the ED deployment distribution controlled by $\kappa\in[-\frac{2}{ R^2} , \frac{2}{ R^2}]$, or the inter-transmission time statistics adjusted by the function $v(\tau)$ described in Sec. \ref{sec:packet}.
Further, the objective function to be optimized may be the network coverage probability $C(\kappa,\lambda_0)$, its MD generalization $\mathcal{C}(\kappa,\lambda_0,z)$, or their per-SF  analogues $C_n(\kappa,\lambda_0)$, $\mathcal{C}_n(\kappa,\lambda_0,z)$ or some statistic of these (e.g., its mean and variance).
In this section, we argue that the appropriate objective function that can capture the great variability between SF rings and thus ensure good and reliable coverage of the region $\V$ is  
$
\mathcal{O}_n(\kappa,\lambda_0, z) =  \mathcal{C}_n(\kappa,\lambda_0,z) \frac{N_n}{|\mathcal{V}_n|}
$
where $|\V_n|=\pi(l_{n}^2 - l_{n-1}^2)$.
Thus, $\mathcal{O}_n(\kappa,\lambda_0, z)$ is a measure of the number of EDs per square km that will achieve a link reliability of $z$ in the $n_\text{th}$ SF ring, given a deployment strategy of $\lambda(d)$ and inter-transmission time statistics $v(\tau)$.
In other words, $\mathcal{O}_n(\kappa,\lambda_0, z)$ is the $z$-\textit{effective} density of EDs.
Further, in order to achieve a fair coverage of the LoRa deployment area $\V$ one should aim to maximize the product $\mathcal{O}(\kappa,\lambda_0, z)=\prod_n \mathcal{O}_n(\kappa,\lambda_0, z) $, or  equivalently the sum $\sum_n \ln \mathcal{O}_n(\kappa,\lambda_0, z) $.
Fig. \ref{fig:optimal} shows the results of such an optimization using $v(\tau)=598\sqrt{\tau}$ and $u=99$. 
A clear optimal deployment strategy is identified at $\kappa=-0.015$ (concave) and $\lambda_0=0.8$ (i.e., $N=\lambda_0 |\V|=293$), resulting in a $z$-effective per-SF density $\mathcal{O}_n$ that ranges from 0.05-0.6 EDs per km${}^2$ with reliability $z=70\%$. 
Intuitively, this solution balances having more(fewer) EDs in SF7(SF12) due to the concave deployment, while being more(less) likely to collide due to the sub-linear spread as seen in the top right of Fig. \ref{fig:packet}. 

\squeezeup
\section{Conclusion \label{sec:conclusion}}

In this letter we have proposed a tractable mathematical framework for the uplink performance analysis of LoRa networks able to model non-uniform deployments. 
While recent investigations have also studied non-uniform settings \cite{blaszczyszyn2019analyzing}, this paper has
derived closed form expressions for the LoRa packet collision probability and coverage probability using meta distribution statistics and demonstrated how the two are intrinsically related through co-SF interference.
Finally we defined optimization problems used to obtain optimal and fair deployment strategies.
We find that concave deployments of LoRa EDs, along with a sub-linear spread of random access inter-transmission times provide for an optimal and fair network coverage.
Multi-gateway generalizations of our results would be an important next step \cite{georgiou2020cov} as well as the validation of our analysis through numerical simulations.

\squeezeup \section{Acknowledgements}
The authors would like to acknowledge funding from the EUs H2020 research and innovation programme under the Marie Sk{\l}odowska-Curie  project NEWSENs, No 787180.
This work was co-funded by the European Regional Development Fund and the Republic of Cyprus through the Research and Innovation Foundation, under the project INFRASTRUCTURES/1216/0017 (IRIDA).

\squeezeup
\bibliographystyle{ieeetran}
\bibliography{mybib}

\end{document}